\documentclass[runningheads]{llncs}
\usepackage{color,soul}

\usepackage{enumitem}
\usepackage[dvipsnames]{xcolor}
\usepackage{graphicx}
\graphicspath{ {./images/} }
\usepackage[cmex10]{amsmath}

\usepackage{algorithm}
\usepackage[noend]{algcompatible}
\usepackage[nodisplayskipstretch]{setspace}
 \usepackage[font=scriptsize,labelfont=bf]{caption}
\begin{document}
\title{
Service-Based  Wireless Energy Crowdsourcing}
%
%
%
%

\author{Amani Abusafia \and
Abdallah Lakhdari \and
Athman Bouguettaya}
\authorrunning{A. Abusafia et al.}
\institute{The University of Sydney, Sydney NSW 2000, Australia\\
\email{\{amani.abusafia,abdallah.lakhdari,athman.bouguettaya\}@sydney.edu.au}}

\maketitle              
\begin{abstract}

We propose a novel service-based ecosystem to crowdsource wireless energy to charge IoT devices. We leverage the service paradigm to abstract wireless energy crowdsourcing from nearby IoT devices as energy services. The proposed energy services ecosystem offers convenient, ubiquitous, and cost-effective power access to charge IoT devices. We discuss the impact of a crowdsourced wireless energy services ecosystem, the building components of the ecosystem, the energy services composition framework, the challenges, and proposed solutions.


\end{abstract}

\section{Introduction}

\textit{Internet of Things (IoT)} is a paradigm that enables everyday objects (i.e., \emph{things}) to connect to the internet and exchange data \cite{whitmore2015internet}.  IoT devices usually have  capabilities, such as sensing, networking, and processing \cite{whitmore2015internet}. The number of connected IoT devices is expected to reach 125 billion in 2030 \cite{markit2017internet}. This potential pervasiveness of IoT provides opportunities to \emph{abstract} their capabilities using the \emph{service paradigm} as \emph{IoT services}\cite{bouguettaya2021internet}. IoT services are defined by their functional and non-functional attributes. The functional attributes define the purpose of the service, such as sharing internet access using WiFi. The non-functional attributes are the properties that assess the Quality of Service (QoS), e.g.,  signal strength, reliability, etc. For example, an IoT device owner may offer their WiFi as a hotspot (i.e., service provider) to other nearby IoT devices (i.e., service consumers). A multitude of novel IoT services may be used to enable intelligent systems in several domains, including smart cities, smart homes, and healthcare \cite{bouguettaya2021internet}. Examples of IoT services are WiFi hotspots \cite{ba2022multi}, environmental sensing \cite{kelly2013towards}, and  energy  services \cite{lakhdari2020composing}. Of particular interest is the use of energy services.\looseness=-1

Energy service, also known as \emph{Energy-as-a-Service (ES)}, refers to the\textit{ wireless power transfer} among nearby IoT devices \cite{lakhdari2020composing}.  We consider a particular set of IoT devices named \textit{wearables}. Wearables refer to anything worn or hand-held like smart shirts, smartwatches, and smartphones \cite{seneviratne2017}. Wearables may harvest energy from natural resources such as kinetic activity, solar power, or body heat \cite{choi2017wearable}\cite{manjarres2021enhancing}. For instance, a smart shoe using a PowerWalk kinetic energy harvester may produce 10-12 watts on-the-move power\footnote{bionic-power.com}. In this respect,  wearing a PowerWalk harvester may generate energy to charge up to four smartphones from an hour's walk at a comfortable speed. Energy services may be deployed through the newly developed ``Over-the-Air" wireless power transfer technologies \cite{lakhdari2020composing}\cite{OvertheAirCharger}.  Several companies focus on developing the wireless charging technology of IoT devices over a distance, including WiTricity\footnote{witricity.com}, Energous\footnote{energous.com}, Cota\footnote{ossia.com}, Powercastco\footnote{powercastco.com}. For example, WiTricity started based on the work of \cite{kurs2007wireless} where they succeeded in transferring 60 W of power wirelessly to power a light bulb. Another example is Energous which  developed a device that can charge up to 3 Watts power within a 5-meter distance to multiple receivers.


 \begin{figure}[!t]
\centering
\includegraphics[width=\linewidth]{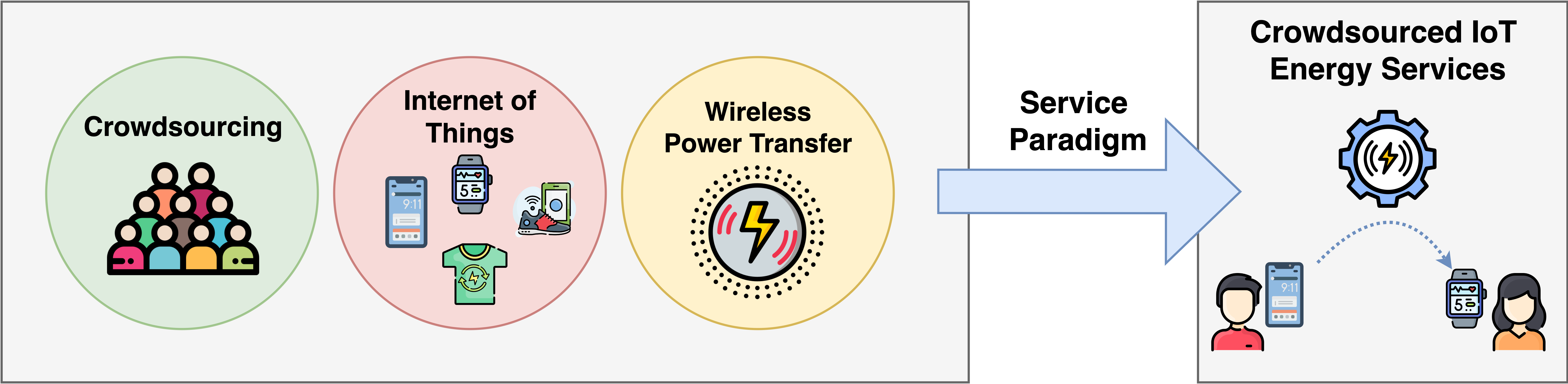}
\caption{The components of  IoT energy services ecosystem}
\label{concept}
\end{figure}



\textit{Crowdsourcing} is an efficient way to  leverage IoT energy services to create a self-sustained environment \cite{lakhdari2020composing}\cite{bouguettaya2021internet}.  IoT users may \textit{collaborate} to share their spare energy to charge nearby IoT devices and extend their battery endurance \cite{fang2018fair}\cite{sakai2021towards}. Crowdsourced IoT energy services present a \textit{convenient} and adaptable solution as devices do not need to be tethered to a power point or use charging cords, and power banks \cite{feng2020advances}. In addition, crowdsourcing energy offers an  \textit{ubiquitous} power access for IoT users as they may be charged anytime and anywhere, even while moving \cite{lakhdari2020composing}\cite{feng2020advances}. Charging IoT devices wirelessly from a central source usually requires a high-frequency magnetic field to transfer the energy over a distance  \cite{lin2013wireless}. Studies have shown that a strong magnetic field has a harmful impact on humans \cite{baikova2016study}\cite{lin2013wireless}. On the contracts, crowdsourcing IoT energy services enables charging by aggregating energy from multiple close-by devices. As the devices are near, transferring the energy will require a low-frequency magnetic field. Hence, crowdsourcing energy services offer an alternative solution to charge devices wirelessly without compromising users' health \cite{Amani2022QoE}\cite{abusafia2022maximizing}. In this paper, we propose to leverage the service paradigm to enable a self-sustained IoT energy services ecosystem by utilizing three components, crowdsourcing, IoT, and wireless power transfer technologies (see Fig.\ref{concept}).

This paper maps out a strategy to leverage the service paradigm to utilize IoT energy services in smart cities. We envision a sustainable ecosystem that allows on-the-go wireless energy crowdsourcing to recharge IoT devices in smart cities. First, we highlight the benefits of the energy services ecosystem. Then, we present a holistic service-based ecosystem to conceptualize the idea and call for future validation. Next, we describe the ecosystem in terms of the environment, service-oriented architecture, and enabling technologies. We also present the envisioned framework and the contemporary approaches for composing IoT-based energy services. Finally, we discuss the uprising challenges to implementing the envisioned ecosystem and highlight the potential future research directions that may address these challenges.



\section{Impact of Crowdsourcing Energy Services Ecosystem }
\label{Benefits}

Crowdsourcing wireless energy services provides numerous benefits to both the environment and IoT users. Each of the benefits is discussed in depth below.
\begin{itemize}
 \setlength\itemsep{0.5em}

\item \textbf{Enabling Sustainable IoT Ecosystem:}
The proliferation of IoT devices leads to a significant increase in energy consumption \cite{sharma2020blockchain}\cite{arshad2017green}. IoT devices' global energy demand is predicted to reach 46TWh by 2025 \cite{arshad2017green}. This tremendous energy demand accounts for 6\% to 8\% of the global carbon footprint generated by information and communication technologies
\cite{shaikh2015enabling} \cite{varjovi2020green}. Crowdsourcing energy services reduce carbon footprint by reducing dependence on fossil fuels to charge IoT devices.   In addition, crowdsourcing energy services rely on renewable or spare energy resources. The renewable energy may be harvested from natural resources such as body heat, solar energy, or kinetic movement \cite{gorlatova2015movers}.

 \begin{figure}[!t]
\centering
\includegraphics[width=0.8\linewidth]{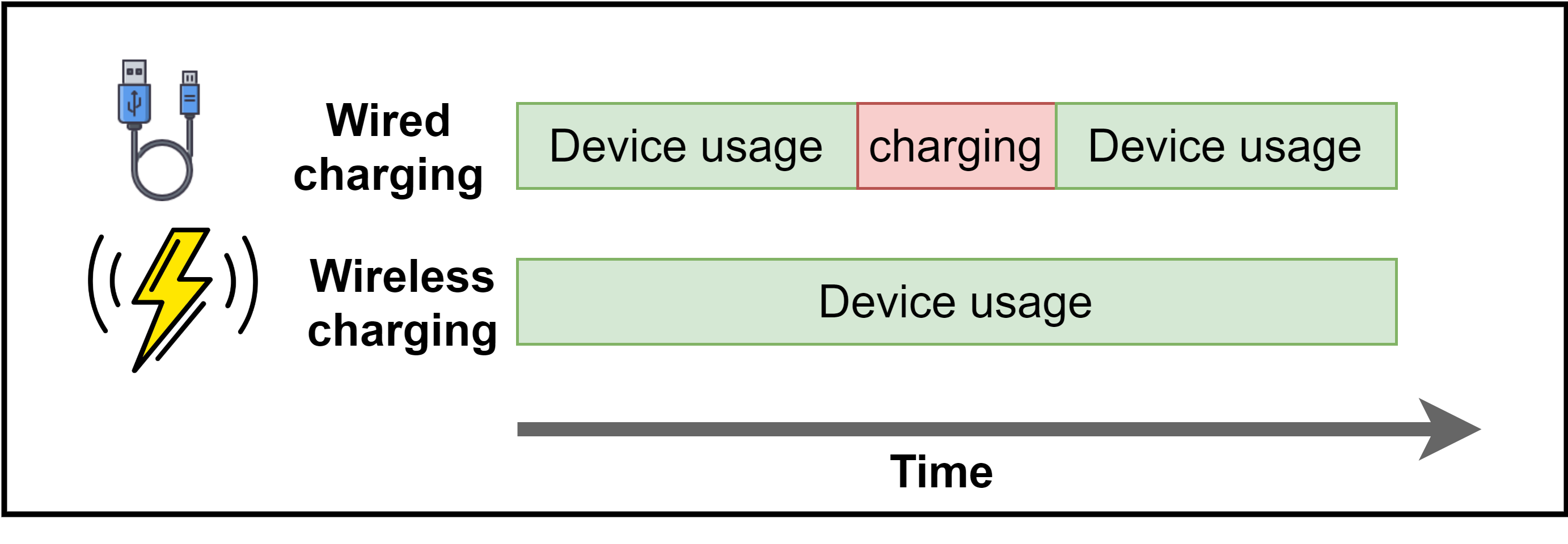}
\caption{Usage-time using wire vs wireless charging technologies}
\label{usage}
\end{figure}
\item \textbf{Extending Battery Endurance:}
IoT devices are constrained by the size of their batteries \cite{pasricha2020survey}\cite{raptis2020wireless}. The small size of the battery results in a limited energy storage capacity. This limited capacity hinders the capabilities of IoT devices. Increasing the battery capacity for IoT devices faces several challenges,including,  safety, weight, cost, and recycling \cite{scrosati2010lithium}. Other charging methods, such as carrying a power cord and searching for a power outlet, are inconvenient for users. Hence, crowdsourcing energy services becomes an attractive solution to improve battery endurance \cite{dhungana2020peer}.
\item  \textbf{Unlimited Usage Time:}
Crowdsourcing energy services enable recharging IoT devices without interrupting their usage time. In contrast, wired charging may require IoT users to stop the usage of their devices in order for them to be charged. Hence, wireless recharging of IoT devices may result in unlimited usage time, i.e., constantly using them to provide or access services (See Fig.\ref{usage}). Thus, crowdsourcing energy services prolong the IoT device's usage lifetime, especially when the external energy supply is unavailable. 

\item \textbf{Spatial Freedom:}
Crowdsourcing wireless energy services provide spatial independence for IoT users. As previously mentioned, IoT devices are frequently required to be tethered to power points or carry power banks to be charged. The last obstacle for IoT devices to achieve their \textit{spatial freedom} is both charging wires and pads. The prospected environment is expected to enable wireless power transfer delivery up to five meters \cite{lakhdari2020crowdsharing}.

\item \textbf{Flexible Contracts: }   Unlike traditional services, crowdsourced energy services environment don't have  lock-in contracts \cite{lakhdari2018crowdsourcing}.  The flexible contracts enable consumers and providers to request or offer full or partial services according to their preferences \cite{lakhdari2018crowdsourcing}\cite{abusafia2020incentive}.
 In addition, providers and consumers may extend their stay-time to offer or receive an energy service \cite{lakhdari2020Elastic}. 
 
 \item \textbf{Ubiquitous Power Access:}
 As aforementioned, energy services offer spatial freedom for IoT devices. Therefore, crowdsourcing energy services offer an \textit{ubiquitous} power access to charge IoT devices anytime and anywhere. This \textit{ubiquity} of wireless charging facilitates access to energy for IoT users.
 
 \item \textbf{Convenience Charging:}
The \textit{ubiquity} of wireless charging offers a convenient alternative to charge IoT devices \cite{feng2020advances}.  In addition, as previously discussed, the \textit{flexibility} to participate in crowdsourcing energy services without any long term  contract offers convenience for IoT users. IoT users might offer or request  energy services without any lock-in contract. They can also move around freely while providing or receiving energy \cite{lakhdari2020fluid}. This convenience ensures that the ecosystem will expand widely.

  
\item \textbf{Business Edge:} 
As previously mentioned, energy services offer a convenient solution to charge IoT devices. Thus, energy services may be used as a complementary service to provide customers with the best quality of experience when visiting a business \cite{Amani2022QoE}. For example,  energy services, like WiFi, may be used to charge customers' wearables in a cafe. Such a service will increase customers' satisfaction with a business. Customer satisfaction is the main factor in ensuring that customers keep coming back to businesses, i.e., maintaining or increasing foot traffic \cite{chao2013c}. For example, a case study showed that "Sacred", a cafe in London, had a noticeable increase in foot traffic after installing wireless charging points\footnote{air-charge.com}. The \textit{foot traffic} in a business has a direct impact on its  revenue \cite{muller1994expanded}. 


	
\end{itemize}

\section{Crowdsourcing Energy Services Ecosystem }

The energy services ecosystem setup consists of three major components: (1) The context of the envisioned prospective environment, (2) The abstraction of the ecosystem as an energy service-oriented architecture, and (3) The enabling technologies to implement the foreseen ecosystem, In what follows, we discuss each component in detail.


 \begin{figure}[!t]
\centering
\includegraphics[width=\linewidth]{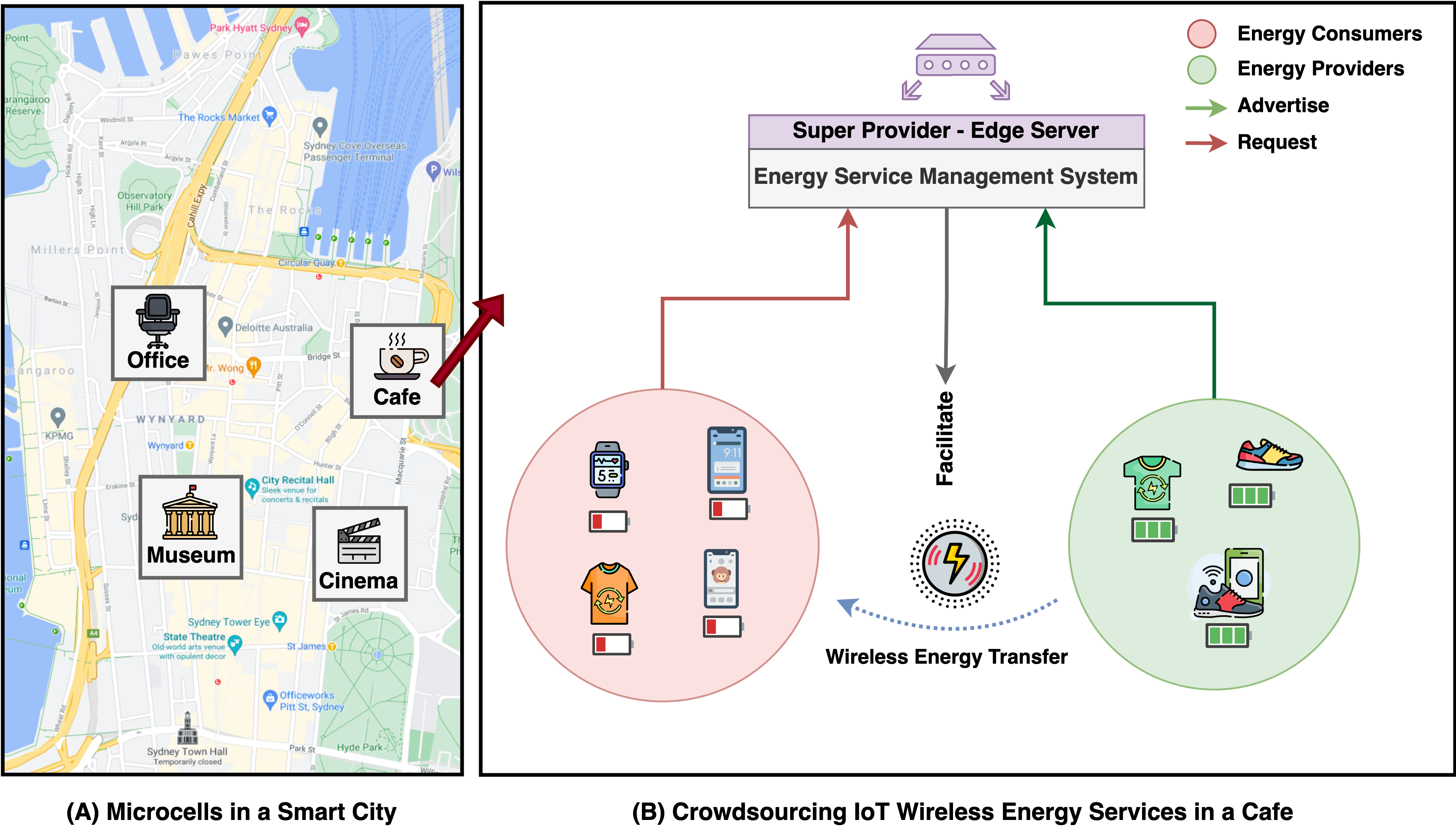}
\caption{Crowdsourcing IoT energy services scenario}
\label{Scenariofig}
\end{figure}

\subsection{Prospective Environment:} The envisioned crowdsourced IoT energy environment is a \textit{dynamic} environment that consists  of IoT users congregating and moving across \textit{microcells} \cite{lakhdari2020composing}\cite{lakhdari2021proactive}. A microcell is any confined area in a smart city where people may gather (e.g., coffee shops, restaurants, museums, libraries) (see Fig.\ref{Scenariofig} (A)). In this environment, IoT users are assumed to act as energy providers or consumers (see Fig.\ref{Scenariofig} (B)).
Their IoT devices are assumed to be equipped with wireless energy transmitters and receivers. Energy providers may use their wearables to harvest energy \cite{TranM0B19}. The IoT users might share their spare or harvested energy to fulfill the requirements of nearby IoT devices. Energy Providers advertise services and consumers submit requests to the IoT coordinator. The super provider manages the IoT coordinator. A super provider is typically a microcell's owner who manages the exchange of energy services between providers and consumers. The IoT coordinator is assumed to be deployed one hop away from the energy providers and consumers (e.g., router at the edge) to minimize the communication overhead and latency while advertising energy services and requests. The super provider limits a consumer's requested energy and uses a reward system to encourage providers to share energy. Rewards come in the form of stored credits to providers. The collected credits may be used later by the provider to increase the limit on the amount of requested energy when they are in the consumer role. Providers may also be a consumer and vice-versa. Providers receive rewards based on the amount of shared energy. 

In the energy services environment, IoT devices may share energy using the energy service model. Note that energy  services  exhibit \emph{functional} and \emph{non-functional} (Quality of service (QoS)) properties \cite{lakhdari2018crowdsourcing}. In this respect, the functional property of energy service is represented by the wireless service delivery of energy to nearby IoT devices. The non-functional (QoS) properties may include the amount of energy, location, duration of sharing, etc.


\subsection{Energy Service Oriented Architecture:} 
\begin{figure}[!t]
\centering
\includegraphics[width=0.75\linewidth]{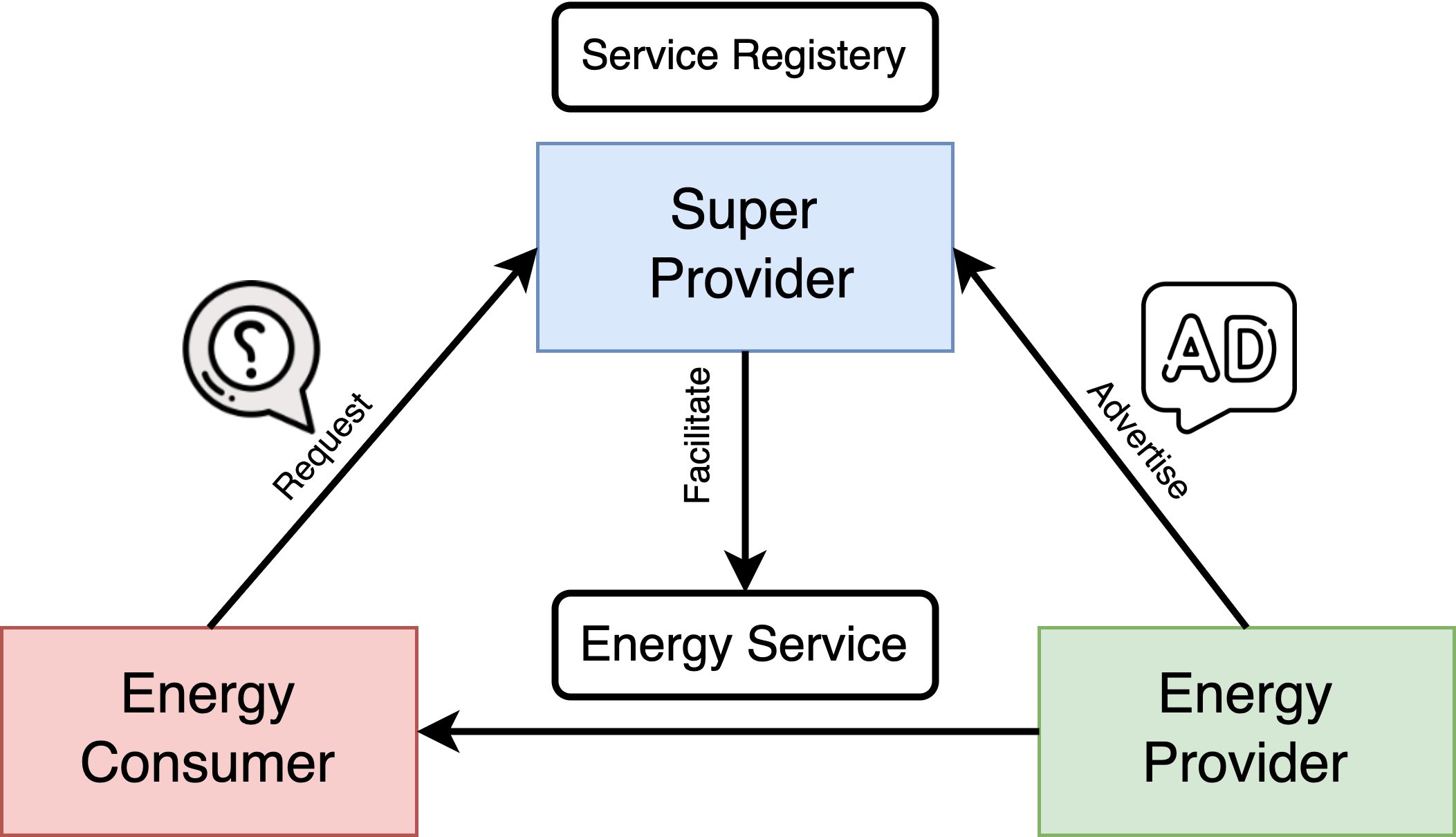}
\caption{ Energy services oriented architecture}
\label{SOA}
\end{figure}

 A key aspect to unlocking the full potential of the energy services ecosystem is to design an \textit{end-to-end} Service Oriented Architecture (SOA) to share crowdsourced energy services. We identify three key components of the SOA: energy service \textit{provider}, energy service \textit{consumer}, and \textit{super provider}. (see Fig.\ref{SOA}). An \emph{energy provider} refers to an IoT device that may share their energy. An \emph{energy consumer} refers to an IoT device that requires energy. A super provider is typically a microcell's owner who manages the exchange of energy services between providers and consumers.

\subsection{Enabling Technologies:}
The implementation of a crowdsourcing energy services ecosystem strongly depends on the recent technology of wireless charging. Wireless charging technologies are the transmitters and receivers that enable the \textit{wireless} transfer of energy \cite{OvertheAirCharger}. For example, Xiaomi’s Mi Air charger transmits energy wirelessly to nearby IoT devices\footnote{mi.com}.

Energy harvesting technologies are another technology to enable a sustainable crowdsourcing energy services ecosystem. Energy harvesters may enable a green crowdsourcing ecosystem to charge IoT devices \cite{lakhdari2021fairness}. The IoT devices may harvest energy from natural resources, including kinetic movement, solar power, or body heat \cite{choi2017wearable}\cite{manjarres2021enhancing}. The harvested energy may be used to charge the device itself or to be shared with nearby devices \cite{worgan2016mobile}.

\section{Crowdsourcing Energy Services Framework}
\begin{figure}[!t]
\centering
\includegraphics[width=\textwidth]{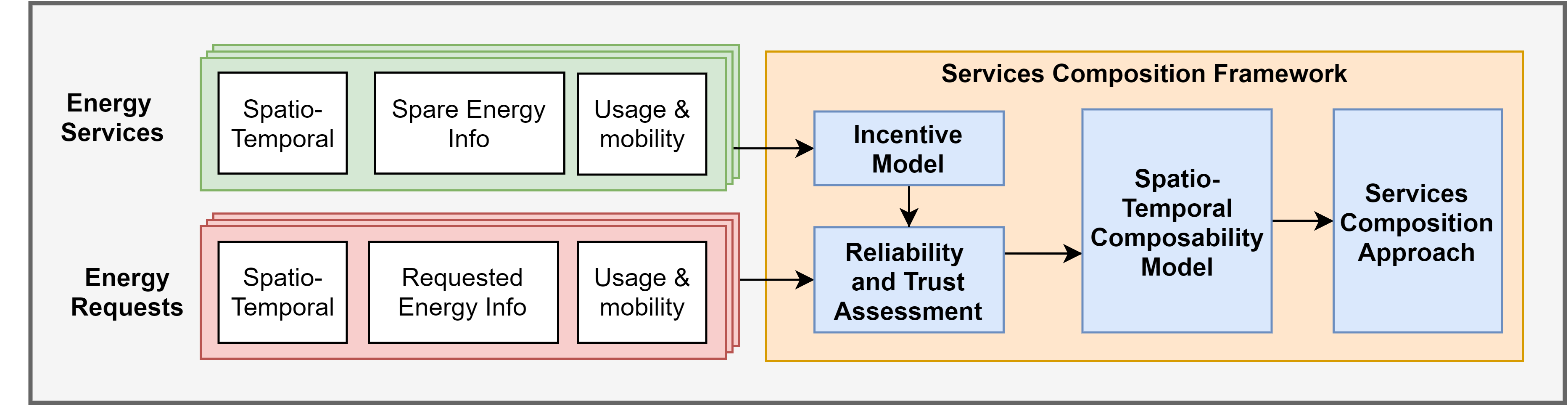}
\caption{Crowdsourcing energy services framework}
\label{framework}
\end{figure}

The energy service framework is responsible for composing and managing the received energy services and requests. An energy service is described by the IoT owner's spatio-temporal preferences, available energy, usage, and mobility model (See Fig.\ref{framework}). Similarly, an energy request is described by the IoT owner's spatio-temporal preferences, amount of requested energy, usage, and mobility model. We envision the composition framework to consist of four components: (1) an incentive model. (2) a reliability and trust assessment, (3) a spatio-temporal composability model, and (4) a service composition approach. In what follows, we discuss each component in detail.
  
\subsection{Incentive Model}
Typically, in crowdsourcing environments, users resist sharing their resources \cite{egger2017crowdsourcing}. Similarly, IoT users may resist sharing their energy since energy is a scarce resource \cite{abusafia2020incentive}\cite{bulut2022special}. Hence, an incentive model is needed to encourage providers to share their energy. Designing an incentive model shall consider the context of the environment and the behavior of the IoT users \cite{wang2022nudging}\cite{abusafia2020incentive}. 

\subsection{Reliability and Trust Assessment}

The participation of IoT users in the energy crowdsourcing ecosystem is influenced by reliable and trustworthy providers, and consumers \cite{abusafia2020Reliability}. To guarantee high-quality services, both reliability and trust are required in crowdsourced IoT systems. High-quality services will encourage and maintain users' participation in the energy services ecosystems. Therefore, when managing and composing energy services, it is imperative to assess the reliability and trustworthiness of IoT energy users. 

\begin{figure}[!t]
\centering
\includegraphics[width=\textwidth]{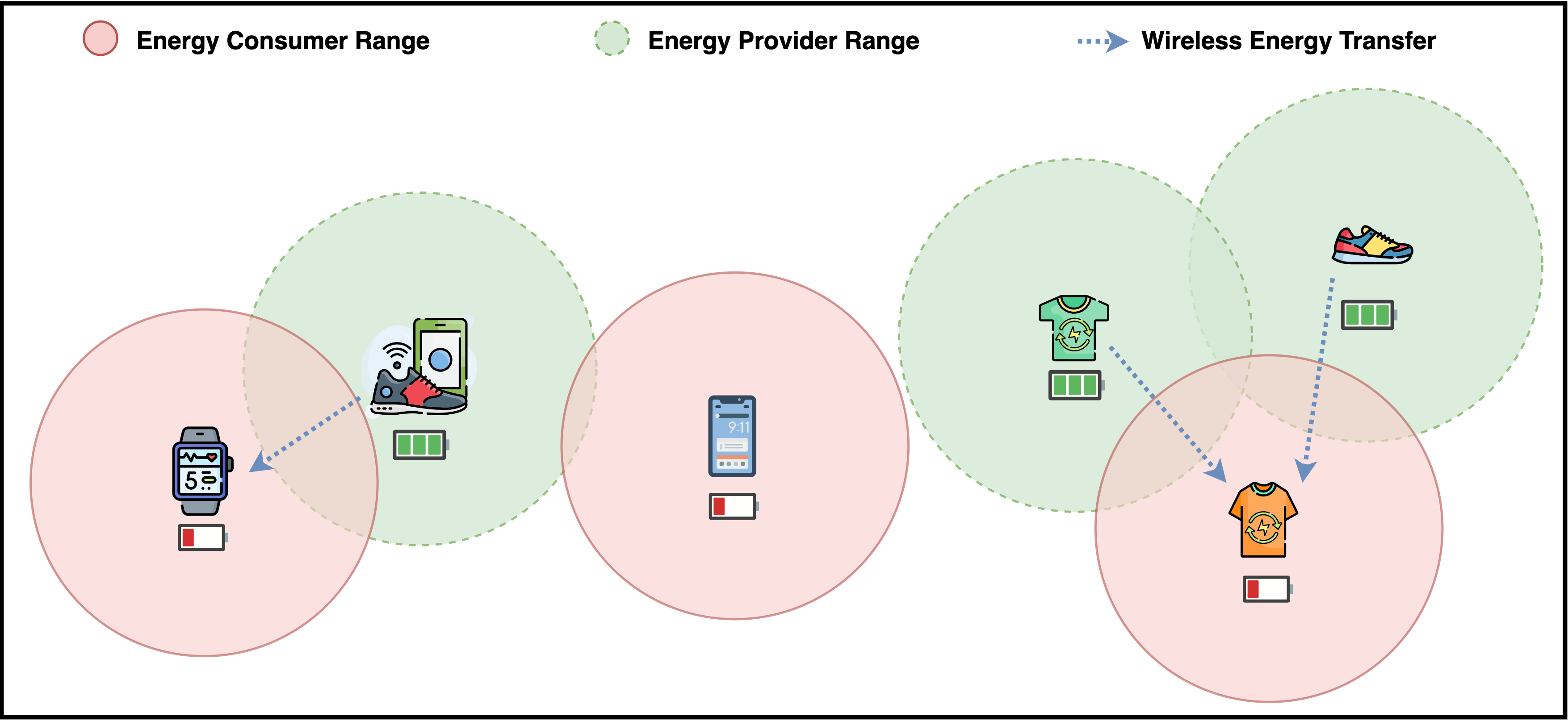}
\caption{Energy consumers may be charged from energy providers within the power transfer range}
\label{powerrange}
\end{figure}

\subsection{Spatio-temporal Composability Model }

  Energy services and requests may have different times and locations \cite{lakhdari2018crowdsourcing}\cite{lakhdari2020composing}. Successfully delivering energy services requires providers and consumers to be within the power transfer range. For example, if a provider's location is out of the power transfer range from a given consumer, the system shouldn't match them (See Fig.\ref{powerrange}). Thus, an effective filtering method is needed in order to efficiently match services to requests. A composability model was proposed to index services nearby a signal request based on time and space \cite{lakhdari2018crowdsourcing}. The same model may be used to filter multiple services and multiple requests.

\subsection{Services Composition Approach}
\begin{figure}[!t]
\centering
\includegraphics[width=\linewidth]{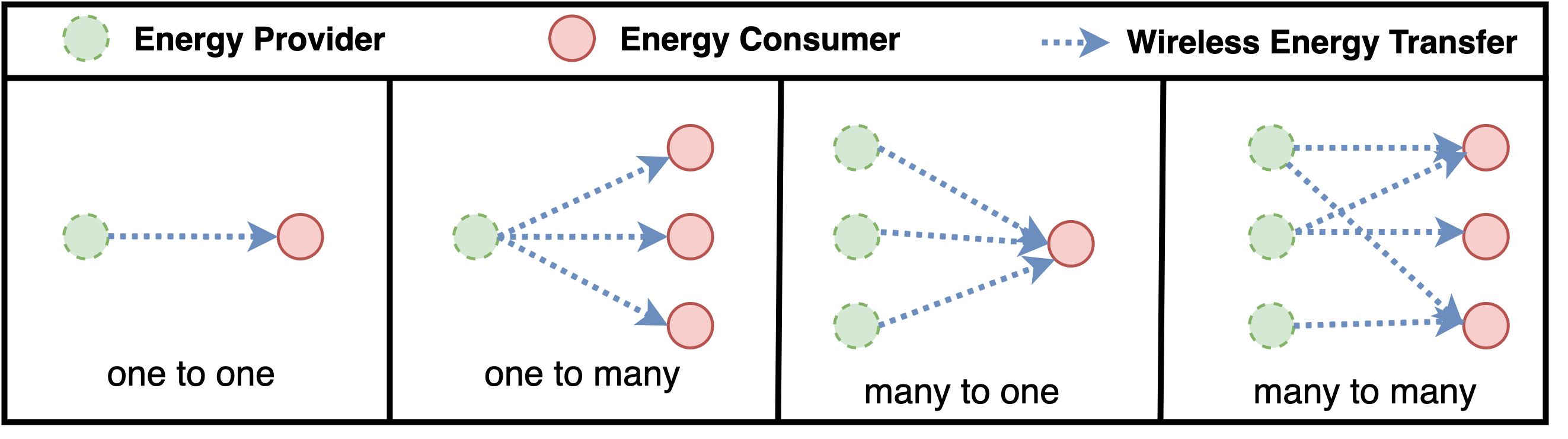}
\caption{Modes of energy services composition}
\label{Comp}
\end{figure}

A single energy request may not be fulfilled by one energy service due to the limited resources of IoT devices \cite{lakhdari2018crowdsourcing}. In such cases, the system may utilize the shareable nature of energy by composing multiple nearby energy services to one request or vice versa \cite{Amani2022QoE}. 
Similarly, a single energy service may be used to fulfill one or multiple energy requests \cite{lakhdari2018crowdsourcing}. Hence, we may have different modes of composition (See Fig.\ref{Comp}). Additionally, energy service providers and consumers may have different Spatio-temporal preferences. Therefore, the service composition approach should consider these preferences in matching and composing energy services and requests.

\section{Challenges in Crowdsourcing Energy Services}
\label{OpenChallenges}
To exploit the potential of crowdsourced energy services for charging IoT devices, multiple challenges must be tackled. These challenges can be classified into three main categories enabling technologies,  human-in-the-Loop, and system deployment. The following subsections describe the challenges encountered in each category and propose possible solutions.

\subsection{\textbf{Enabling Technologies:}}
The deployment of a crowdsourcing energy services ecosystem highly relies on the current Wireless Power Transfer (WPT) technologies. Although WPT technologies are advancing rapidly, there are technological constraints that impede the ecosystem from being fully deployed \cite{WirelessMarket2027}. Challenges posed by technology include energy loss, miniature energy delivery, and restricted power transmission range. Each of these constraints is discussed further below.

\subsubsection{\textbf{Energy Loss}} Using the recently developed wireless power transfer technologies, consumers may spend more energy to receive the requested energy \cite{yao2022wireless}. The consumed energy may include the energy required for service discovery, energy delivery, and connection establishment between the consumer and the provider \cite{na2017energy}.
The huge energy loss impedes the energy services ecosystem from fulfilling its purpose of charging nearby IoT devices. Therefore, minimizing energy loss is a challenge yet to be addressed. 

\subsubsection {\textbf{Miniature Delivered Energy}}
Even though there is rapid development in wireless power transfer technologies, the amount of delivered energy is insufficient to what is currently required \cite{feng2020advances}. The small amount of delivered energy may be justified by the energy loss in delivering the energy service as aforementioned. Hence, Delivering a considerable amount of energy wirelessly over a distance is still a challenge that hinders the full deployment of the energy services ecosystem.


\subsubsection {\textbf{Limited Power Transfer Range}}

The near-field magnetic coupling is an efficient technology to transfer energy wirelessly \cite{feng2020advances}. However, the magnetic coupling has a short power transfer distance. The limited transfer distance restricts the spatial freedom we envision as part of this ecosystem. 

\subsection{Human-in-the-loop}

Crowdsourcing IoT energy services is a human-centered application as it depends on their participation behavior. Modeling human behavior is challenging as it requires modeling complex behavioral, psychological and physiological characteristics of human nature \cite{delicato2020challenges}. In our environment, the human factor has several influences on the ecosystem. First, the deployment of the ecosystem depends on the IoT users' willingness to participate in the energy sharing process \cite{abusafia2020incentive}. Second, the mobility of IoT users in this dynamic environment could impede the delivery of energy services \cite{lakhdari2020fluid}. Third,  energy services are offered by IoT devices that are simultaneously in use by their owners. Therefore, the consistent wireless energy delivery from one IoT device to another depends on the device owners' usage frequency. Finally, the spatial and temporal preferences of IoT users differ, which affects the availability of energy services. In what follows, we discuss each of the aforementioned challenges.

\subsubsection{\textbf{Provisioning Resistance }}

 Individuals are likely motivated to engage in the energy services ecosystem since they can easily access nearby energy resources. However, since energy is a scarce resource, they may be reluctant to share it  \cite{abusafia2020incentive}. Service resistance is the reluctance to provide services due to limited resources \cite{deng2016toward}. In the energy services context, service resistance refers to the unwillingness to provide energy to IoT devices \cite{abusafia2020incentive}. A provider's resistance is influenced by the provider's available energy and the size of the requested energy.   As a result, one of the challenges is to predict providers' resistance to offering their energy service and devise a correct incentive to overcome it. Since humans get impacted differently by different incentives, determining the effective incentive for energy providers is a challenge that still needs to be addressed.

\subsubsection{\textbf{Modelling Human Mobility}}
IoT devices change their locations regularly according to the mobility of their owners. The mobility of the crowd has multiple impacts on the crowdsourcing ecosystem. First, the mobility of the crowd across microcells determines the availability of energy services and requests. Second,
 The IoT users' mobility affects the connectivity between IoT devices. Highly moving IoT devices within a microcell may result in disconnecting the wireless energy transfer between them. A disconnection in the delivery of the energy transfer may result in its failure \cite{abusafia2020Reliability}\cite{lakhdari2020fluid}. Therefore, the mobility of  IoT users should be modeled and analyzed to manage the energy services in the ecosystem better. The mobility model of  IoT users should be studied and investigated on a micro level as indoor mobility and a macro level as mobility patterns across microcells. The indoor mobility models should reflect the users' movement within a confined space \cite{menon2022diy}. Indoor mobility models will enable the ecosystem to filter services and requests from a spatial perspective within the power transfer range. Moreover, it will allow the ecosystem to predict users' movement and their impact on disconnecting a service delivery. On the other side, the outdoor mobility pattern of an IoT user may be used to proactively plan for them when to request energy \cite{lakhdari2021proactive}. Furthermore, it enables us to understand the crowd movement and predict the undersupplied areas \cite{zheng2022imap}. Thus, mobility models are needed to unlock the full potential of the ecosystem.
 \subsubsection{\textbf{Modelling Usage Patterns}}: 
 The energy usage pattern of providers may impact the consistency of provisioning energy services \cite{lakhdari2020Elastic}. 
 An energy provider with highly dynamic usage of their device impacts the delivered energy service quality (QoS). Providers may offer their spare energy while simultaneously using it for their devices. If a provider is heavily using their device, they may end up consuming their advertised service while sharing it. Thus, energy services may fluctuate due to the providers' usage.   As a result, IoT devices' energy usage behavior must be modeled to ensure consistent provision of energy services \cite{menon2022diy}\cite{zheng2022imap}.\looseness=-1 

\subsubsection {\textbf{Determining Services Availability:}} 
Matching energy services and requests in a crowdsourced environment is challenging due to the uncertain availability of energy services and requests \cite{Amani2022QoE}. Predicting the availability of energy services is challenging as it relies on detecting the mobility patterns of IoT Users and their energy usage patterns. Defining the mobility patterns mainly focuses on analyzing the spatio-temporal attributes and potential regularities hidden in individual movement trajectories \cite{wang2019urban}. However, the existing literature to determine mobility patterns has low accuracy due to the \textit{flexible nature} of human movement \cite{zheng2022imap}. Thus, the challenge of detecting human mobility patterns hinders the prediction of service availability.

\subsection{\textbf{Deployment of Energy Services Ecosystem}}

The deployment of the energy crowdsourcing ecosystem raises several challenges. In what follows, we discuss each challenge and its impact of  on the  ecosystem.   

\subsubsection{Balancing Local Provisioning and Demand:}
Naturally, energy users' preferences to provide or receive energy vary in time and space. The variety in energy users' preferences may severely affect the \textit{balance} between the amount of required and provided energy within a confined area. The \textit{imbalance} between the energy \textit{demand} and the available services  may result in \textit{oversupplied} and \textit{undersupplied} areas in a smart city. Under-supplied areas will result in unfulfilled energy requests, which may discourage the participation of IoT users in the ecosystem. Hence, ensuring the availability of energy services to fulfill the requirement of consumers is essential to the sustainability of the ecosystem. A redistribution of energy service across over-supplied and under-supplied areas may achieve better-balanced energy availability. Another possible direction is using an incentive model to increase participation in provisioning energy services \cite{abusafia2020incentive}.

\subsubsection{Loose Contracts}

 As previously mentioned, energy services offer flexibility as they do not require lock-in contracts \cite{lakhdari2020Elastic}. For instance, providers may offer full or partial services according to their preferences \cite{lakhdari2018crowdsourcing}. However, this flexibility may impact the commitment of IoT users. The uncertain commitment of IoT users may add uncertainty to the energy transfer process and thereby impact the participation of IoT users in the ecosystem.
    
\subsubsection {\textbf{Reliability}}
Considering reliability in the energy ecosystem encourages the participation of energy users in crowdsourcing energy services \cite{abusafia2020Reliability}\cite{lakhdari2020composing}. The reliability of an energy provider refers to the probability that an energy service will be successfully delivered with the advertised Quality of Service (QoS) attributes. Similarly, the reliability of an energy consumer refers to the probability that an IoT energy service will be successfully received with the same submitted requirements. The reliability of the energy provider is impacted by their usage pattern and mobility. As aforementioned, a provider that heavily uses their device might consume their advertised energy. Also,  a highly mobile provider will cause frequent disconnections in the energy transfer between the devices—similarly, the reliability of an energy consumer is impacted by their mobility patterns. Like providers, a highly mobile consumer will cause frequent disconnections in the energy transfer between the devices. Accordingly, it is vital to consider the reliability of energy users while managing and composings energy services.

\subsubsection {\textbf{Trust}}
The definition of trust is "the confidence, belief, and expectation regarding the reliability, integrity, and other characteristics of an entity"\cite{yan2014survey}. To guarantee high-quality services, trust assessment is required in crowdsourced IoT systems \cite{ba2022multi}. High-quality services will encourage and maintain users' participation in the energy services ecosystems. 
The trust assessment must evaluate the trustworthiness of energy providers and consumers. The trustworthiness of energy providers may represent the provider's reputation, their service reliability, and their security level \cite{ba2022multi}. Likewise, energy consumers' trustworthiness may represent their reputation and their security level. Although trust assessment has been studied in other areas, it remains a challenge in IoT ecosystems due to several reasons \cite{sha2018security}. First, the dynamic nature of the crowdsourced IoT environment makes it challenging to keep an accurate record of the devices' reputations. For instance, IoT devices are usually moving, and their existence may be temporary. Second, IoT devices typically lack a global identifier, which makes it difficult to maintain a globally accessible profile for these devices. As a result, a novel trust assessment framework is required to assess the reputation of IoT devices \cite{sha2018security}.
\subsubsection {\textbf{Security}}
The goal of IoT device security is to prevent and protect against IoT attacks and service failures \cite{lu2018internet}. 
Attacks on IoT devices may compromise the privacy and confidentiality of users, infrastructures, data, and IoT devices \cite{hassan2019current}. Furthermore, IoT attacks such as denial-of-service attacks may impede the delivery of services to IoT Users. Securing IoT systems is a critical challenges because users may not adopt many IoT systems if they are not secure \cite{liyanage2020iot}.
Existing security architectures and protocols are difficult to integrate in IoT devices due to their limited computing power and storage size\cite{payton2018envisioning}. 

\subsubsection {\textbf{Privacy}}
The global digital data generated by IoT devices is expected to reach 180 Zettabytes by 2020 \cite{kanellos_2016}. 
As a result, protecting IoT data and the privacy of users who generate or consume data has become a major concern in research and industry \cite{mendez2018internet}. Privacy refers to ``The claim of individuals, groups, or institutions to determine for themselves when, how and to what extent information about them is communicated to others'' \cite{westin1967privacy}. Maintaining the privacy in IoT ecosystems is challenging as it conflict with the need to use the IoT data to achieve their functions. The data utilization may invade the privacy of IoT users \cite{sha2018security}.
 For example, IoT Data are required to enhance the  energy crowdsourcing process by profiling the energy providers and consumers. The collected IoT data may include the IoT users' mobility behavior and their energy consumption model. Collecting such personal information might violate the privacy of the IoT users. Therefore, maintaining a balance between IoT data protection and use is a continuing challenge.

\section{Conclusion}
\label{conclusion}
We presented a novel service-based framework to crowdsource wireless energy from neighboring IoT devices to charge low-in-battery IoT devices. First, we highlighted the benefits of adopting a crowdsourced wireless energy services ecosystem. We then envisioned and designed an architecture to implement the proposed ecosystem. Lastly, we discussed the open challenges and recommended research directions for possible solutions.

\section*{Acknowledgment}
This research was partly made possible by  LE220100078 and LE180100158 grants from the Australian Research Council. The statements made herein are solely the responsibility of the authors.
\bibliographystyle{unsrt}
\bibliography{main}

\end{document}